\documentclass[a4paper,twocolumn,twoside,10pt]{article}
\usepackage{graphicx,amsmath,amssymb,amsthm,subfig,soul,bbm}
\usepackage{ragged2e}

\usepackage{fancyhdr}
\makeatletter \oddsidemargin-.25in \evensidemargin-.25in
\makeatother \topmargin-0.75in \textwidth7in \textheight9.5in
\begin{document}
\setcounter{page}{51}
\date{}

\renewcommand{\headrulewidth}{0.0pt}
\renewcommand{\footrulewidth}{0.0pt}

\fancypagestyle{all}{
\fancyhead[RO,RE]{}
\fancyhead[LO,LE]{}
\fancyhead[CO]{$~$ \hfill
{\rm A Hybrid Network/Grid Model of Urban Morphogenesis and Optimization} \hfill$~$}
\fancyhead[CE]{$~$ \hfill {\rm J. Raimbault, A. Banos \& R. Doursat} \hfill $~$}
\fancyfoot[RO]{\hspace{1cm}\thepage}
\fancyfoot[RE]{ICCSA 2014, Normandie University, Le Havre, France – June 23-26, 2014}
\fancyfoot[LO]{ICCSA 2014, Normandie University, Le Havre, France – June 23-26, 2014}
\fancyfoot[LE]{\thepage}
\fancyfoot[CO,CE] {}
}

\fancypagestyle{plain}{
\fancyhf{}
\fancyhead[L]{\justify\footnotesize Raimbault, J., Banos, A. \& Doursat, R. (2014) A hybrid network/grid model of urban morphogenesis and optimization. Proceedings of the 4th International Conference on Complex Systems and Applications (ICCSA 2014), June 23-26, 2014, Université de Normandie, Le Havre, France; M. A. Aziz-Alaoui, C. Bertelle, X. Z. Liu, D. Olivier, eds.: pp. 51-60.}
\fancyfoot[L]{ICCSA 2014, Normandie University, Le Havre, France – June 23-26, 2014}
\fancyfoot[R]{\thepage}
}


\title{\bigskip\begin{flushleft}
\noindent {\small {\it Proceedings of ICCSA 2014
\\Normandie University, Le Havre, France - June 23-26, 2014\\[5mm]}}
\end{flushleft}
\Large\bf A Hybrid Network/Grid Model of\\Urban Morphogenesis and Optimization}

\author{Juste Raimbault
\thanks{Graduate School, \'Ecole Polytechnique, Palaiseau, France; and LVMT, Ecole Nationale des Ponts et Chauss\'ees, Paris, France. \mbox{E-mail}: \mbox{juste.raimbault@polytechnique.edu}},
Arnaud Banos
\thanks{G\'eographie-cit\'es, CNRS UMR8504, Paris, France. \mbox{E-mail}: \mbox{arnaud.banos@parisgeo.cnrs.fr}},
and Ren\'e Doursat
\thanks{Complex Systems Institute, Paris \^Ile-de-France (ISC-PIF), CNRS UPS3611. \mbox{E-mail}: \mbox{rene.doursat@iscpif.fr}}\\[10mm]}

\maketitle

\pagestyle{all}


{\small \noindent {\bf Abstract.} We describe a hybrid agent-based model and simulation of urban morphogenesis. It consists of a cellular automata grid coupled to a dynamic network topology. The inherently \emph{heterogeneous} properties of urban structure and function are taken into account in the dynamics of the system. We propose various layout and performance measures to categorize and explore the generated configurations. An \emph{economic evaluation} metric was also designed using the sensitivity of segregation models to spatial configuration. Our model is applied to a real-world case, offering a means to optimize the distribution of activities in a zoning context.}

\section{Introduction} \label{sec_intro}

Recent progress in many disciplines related to urban planning can be interpreted as the rise of a ``new urban science'' according to Batty~\cite{batty2013new}. From agent-based models in quantitative geography~\cite{heppenstall2012agent}, in particular the successful Simpop series by Pumain et al.~\cite{pumain2012multi}, to other approaches termed ``complexity theories of cities'' by Portugali~\cite{portugali2012book}, involving physicists of information theory such as Haken~\cite{haken2003face} or architects of ``space syntax theory'' such as Hillier~\cite{hillier1976space}, the field is very broad and diverse. Yet, all these works share the view that urban systems are quintessentially \emph{complex systems}, i.e.~large sets of elements interacting locally with one another and the environment, and collectively creating a emergent structure and behavior. Taking into account the intrinsic \emph{heterogeneity} of geographical and urban systems, this view lends itself naturally to an agent-based modeling and simulation (ABMS) approach.

Among the most popular ABMS methods are cellular automata (CA), in which agents are cells that have fixed locations on a grid and evolve according to the state of their neighbors. CA models of urban planning, in particular the reproduction of existing urban forms and land-use patterns, have been widely studied, notably by White and Engelen~\cite{white1993cellular}, then analyzed~\cite{batty1997cellular,batty1997possible} and synthesized~\cite{Bat07} by Batty. A recent review by Iltanen~\cite{iltanen2012cellular} of CA in urban spatial modeling shows a great variety of possible system types and applications. They include, for example, ``microeconomic'' CA for the simulation of urban sprawl~\cite{DBM11}, ``linguistic'' CA (including real-time rule update via feedback from the population) for the measure of sustainable development in a fast growing region of China~\cite{Wu96alinguistic}, and one-dimensional CA~\cite{peeters2009space} showing discontinuities and strong path-dependence in settlement patterns.

In this context, we propose a \emph{hybrid} model of urban growth that combines a CA approach with a graph topology containing long-range edges. It is inspired by Moreno et al.'s work~\cite{MBB09,moreno2007conception}, which integrates a network dynamics in a CA model of urban morphology. Its goal was to test the effects of physical proximity on urban development by introducing urban mobility in a network whose evolution was coupled with the evolution of urban shape. We generalize this type of model to take into account \emph{heterogeneous urban activities} and the \emph{functional properties} that they create in the urban environment. This idea was introduced by White~\cite{white2006modeling} and explored by van Vliet et al.~\cite{van2012activity} but, to our knowledge, never considered from the perspective of \emph{physical accessibility} and its impact on sprawl patterns.

The rest of the paper is organized as follows. The model and indicator functions used to quantify the generated patterns are explained in Section~\ref{sec_model}. Next, Section~\ref{sec_results} presents the results of internal and external validations of the model by sensitivity analysis and reproduction of typical urban patterns. It is followed by an application to a concrete case, proposing a bi-objective optimization heuristic of functional configuration based on the relevant objective functions from the validation study. We end with a discussion and conclusion in Sections~\ref{sec_discussion} and~\ref{sec_conclusion}.

\section{Model description} \label{sec_model}

\subsection{Agents and rules}

\begin{figure}[t]
\centering
\includegraphics[width=0.9\columnwidth]{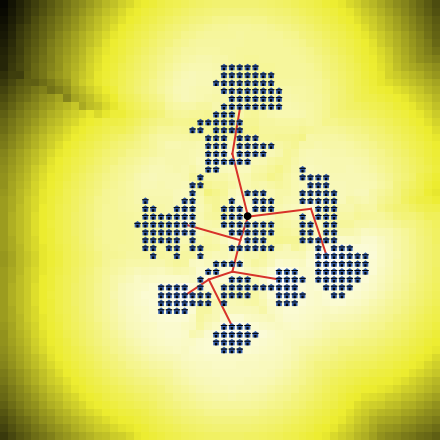}
\caption{\small Example of urban morphology generated by the model. Houses (blue squares) were built in some cells of a $56\times 56$ lattice. City centers and roads (red edges) compose the added network. Cell shades (yellow) represent distances to the built cells (the brighter, the closer).}
\label{fig_lattice}
\vspace{0.5cm}
\includegraphics[width=0.9\columnwidth]{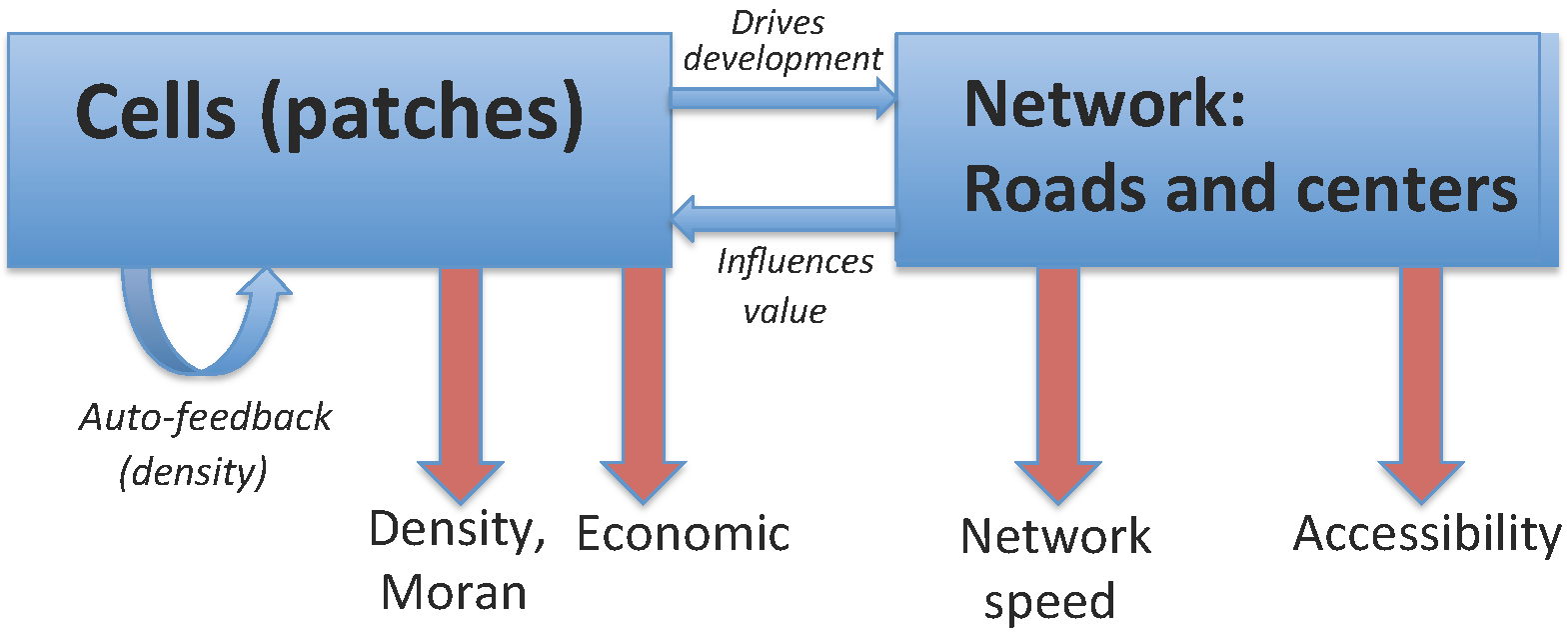}
\caption{\small The hybrid network/grid model. \emph{Blue arrows}: feedback interactions. \emph{Red arrows}: output evaluation functions.}
\label{fig_flowchart}
\end{figure}

The world is represented by a square lattice $(L_{i,j})_{1\leq i,j\leq N}$ composed of cells that are empty or occupied (Fig.~\ref{fig_lattice}). This is denoted by a function $\delta(i,j,t)\in\{0,1\}$, where time $t$ follows an iterative sequence $t\in\mathbb{T} = \tau\mathbb{N} = \{0, \tau, 2\tau, ...\}$~\cite{golden2012modeling} with a regular time step $\tau$. Another evolving structure is laid out on top of the lattice: a Euclidean network $G(t)=(V(t),E(t))$ whose vertices $V$ are a finite subset of the world and edges $E$ (its agents) represent \emph{roads}. In the beginning, the lattice is empty: $\delta(i,j,0)=0$, and the network is either initialized randomly (e.g.~uniformly) or set to a user-specified configuration $G(0)=(V_0,E_0)$. In order to translate functional mechanisms into the growth of a city, we assume that the initial vertices include a subset formed by \emph{city centers}, $C_0\subset V_0$, which have integer \emph{activities}, denoted by $a:C_0\rightarrow\{1,\ldots,a_{\max}\}$.

To characterize the urban structures emerging in this world, we define in general a set of $k$ functions of the lattice, $(d_k(i,j,t))_{1\leq k\leq K}$, called \emph{explicative variables}. These variables are here: $d_1$, the \emph{density}, i.e.~the average $\delta$ around a cell $(i,j)$ in a circular neighborhood of radius $\rho$; $d_2$, the Euclidean \emph{distance} of a cell to the nearest road; $d_3$ the \emph{network-distance} of a cell to the nearest city center, i.e.~the sum of $d_2$ and edge lengths; and $d_4$, the \emph{accessibility} of activities (or rather difficulty thereof), written
\begin{equation}
d_4(i,j,t)=\left(\frac{1}{a_{\max}}\sum_{a=1}^{a_{\max}}d_3(i,j,t;a)^{p_4}\right)^{1/p_4}
\end{equation}
where $d_3(i,j,t;a)$ is the network-distance of the cell to the nearest center with an activity $a$, and $p_4\geq1$ (typically~3) defines a $p$-norm.

A set of weights $(\alpha_k)_{1\leq k\leq K}\in[0,1]^K$ is assigned to these variables to tune their respective influence on what we define as the net \emph{land value} of a cell, as follows:
\begin{equation}
v(i,j,t)=\frac{1}{\sum_k \alpha_k}\sum_{k=1}^K \alpha_k\;\frac{d_{k,\max}(t)-d_k(i,j,t)}{d_{k,\max}(t)-d_{k,\min}(t)}.
\end{equation}

\noindent Houses are preferentially built where $v$ is high, i.e.~$d_k$'s are low. Thus the evolution of the system proceeds in three phases at each time step: (a)~all values $v(i,j)$ are updated, (b)~among the cells that have the best values, $n$ new cells are randomly chosen and ``built'' (set to $\delta=1$); (c)~for each built cell, if $d_2$ is greater than a threshold $\theta_2$ (maximum isolation distance), then that cell is directly connected to the network by creating a new road branching out orthogonally from the nearest edge.

Network initialization is random (see details in~\ref{sec_extval}), and the selection of new cells is also random among identical values of~$v$. A sensitivity analysis and model exploration is conducted in the next section to determine the relative effect of parameters with respect to these sources of randomness. In any case, growth is halted after a constant amount of time $T$, evaluated from experiments, so that the final structure is neither ``unfinished'' nor filling out the world (see~\ref{sec_extval}). Fig.~\ref{fig_flowchart} displays the core ABMS flowchart with feedback interactions between agents.

\subsection{Evaluation functions}

Once a structure is generated, its properties need to be quantified so that it can be categorized or compared to other structures for optimization purposes. To this goal, we define various \emph{evaluation functions}, both objective quantification measures and structural fitness values. The measures described in this section take into account all the explicative variables, whose distributions over the grid are emergent properties that cannot be known in advance and are therefore essential to monitor.

\paragraph{Morphology}

To assess the morphological structure of an urban configuration, we map it onto a 2D metric space defined by a pair of global indicators $(D,I)$ called the \emph{integrated density} and the \emph{Moran index} (Fig.~\ref{fig_morpho}). The density $D\in[0,1]$ is calculated by taking the $p$-norm (with exponent $p_D\geq 1$, typically 3) of the local densities $d_1$:
\begin{equation}
D(t)=\left(\frac{1}{\sum_{i,j} \delta(i,j,t)}\!\!\sum_{\scriptstyle i,j=1\atop\scriptstyle\delta(i,j,t)\neq0}^N\!\!d_1(i,j,t)^{p_D}\right)^{1/p_D}
\end{equation}
\noindent Moran's $I$, an index of spatial autocorrelation, is widely used in quantitative geography~\cite{tsai2005quantifying,lenechet:hal-00696445} to evaluate the ``polycentric'' character of a distribution of populated cells. It is defined by
\begin{equation}
I(t)=\frac{M^2}{\sum_{\mu\neq\nu} 1/d_{\mu\nu}}\frac{\sum_{\mu\neq\nu} (P_\mu-\overline{P})(P_\nu-\overline{P})/d_{\mu\nu}}{\sum_{\mu=1}^{M^2}(P_\mu-\overline{P})^2}
\end{equation}
where the lattice is partitioned into $M\times M$ square areas, at an intermediate scale between cell size and world size ($1\ll\!M\ll\!N$), $d_{\mu\nu}$ is the distance between the centroids of areas $\mu$ and $\nu$, $(P_\mu)_{1\leq\mu\leq M^2}$ denotes the number of occupied cells in each area, and $\overline{P}$ is their global average. We can recognize in this formula the normalized ratio of a modified covariance (pairwise correlations divided by distances) and the variance of the distribution. Moran's $I$ belongs by construction to the interval $[-1,1]$, where values near 1 correspond to a strong monocentric distribution, values around 0 to a random distribution, and values near $-1$ to a checkered pattern (every other cell occupied). Usually, polycentric distributions have relatively small positive $I$ values, depending on the size and distance between centers.

\paragraph{Network performance}

Due to the branching nature of the growth algorithm, the network of roads $G$ cannot contain any other loops than the ones initially present in $G_0$. Therefore, notions of ``clustering coefficient'' or ``robustness'' (with respect to node removal) are not relevant here. On the other hand, since $G$ is intended to simulate a \emph{mobility} network, we can evaluate its performance by defining a \emph{relative speed}~\cite{banos2012towards} $S$, representing the ``detours'' imposed by $G$ with respect to direct, straight travels:
\begin{equation}
S(t)=\left(\frac{1}{\sum_{i,j}\delta(i,j,t)}\!\!\sum_{\scriptstyle i,j=1\atop\scriptstyle\delta(i,j,t)\neq0}^N\!\!\left(\frac{d_3(i,j,t)}{e_3(i,j,t)}\right)^{p_S}\right)^{1/p_S}
\end{equation}
\noindent where $p_S\geq 1$ (also 3), and $e_3(i,j,t)$ is the direct Euclidean distance between cell $(i,j)$ and the nearest city center over the network, i.e.~the one that realizes the value of $d_3(i,j,t)$. Note that $S\ge1$ and is actually higher for more convoluted networks (thus it is a measure of ``slowness'', but we still employ ``speed'').

\paragraph{Functional accessibility}

The global functional accessibility $A$ to city centers is another $p$-norm (also 3), based on the relative local accessibility from each cell, which is $d_4$ over its maximum:
\begin{equation}
A(t)=\left(\frac{1}{\sum_{i,j}\delta(i,j,t)}\!\!\sum_{\scriptstyle i,j=1\atop\scriptstyle\delta(i,j,t)\neq0}^N\!\!\left(\frac{d_4(i,j,t)}{d_{4,\max}(t)}\right)^{p_A}\right)^{1/p_A}
\end{equation}

\noindent This normalization puts $A$ in $[0,1]$ and allows comparing configurations of different sizes. Like $S$, ``better'' urban configurations are characterized by a lower $A$.

\paragraph{Economic performance}

It was shown by Banos~\cite{banos2012network} that the Schelling segregation model, a standard ABM of socio-economic dynamics~\cite{schelling1969models}, was highly sensitive to the spatial structures in which it could be embedded, since segregation rules depended on proximity. This justifies the use of this model as an evaluation of \emph{economic performance} of our urban configurations, measuring how much structure influences segregation. To this aim, we implemented a model of residential dynamics based on the work of Benenson et al.~\cite{benenson1998multi}. The output function is a segregation index $H(t)$ calculated on the residential patterns that emerge inside a distribution of built patches. For urban structures produced in a practical case (Section~\ref{sec_practapp}), we obtained densities of mobile agents between 0.1 and 0.2. Following Gauvin et al.~\cite{gauvin2009phase}, the phase diagram of the Schelling model indicates that in such a density range, tolerance thresholds of 0.4 to 0.8 lead to clustered frozen states, where the calculation of a spatial segregation index is indeed relevant. The detailed description of this economic model is out of the scope of this paper.

\section{Results} \label{sec_results}

Our hybrid network/grid model was implemented in NetLogo~\cite{NetLogo}. Plots and charts were created in R~\cite{R} from exported data. Processing of GIS Data (for vectorization by hand of simple raster data) was done in QGIS~\cite{QGIS_software}. Exploration of the 4D space of explicative variables' weights $\alpha_k$ was conducted inside the $[0,1]^4$ hypercube with a linear increment of 0.2. This created $6^4-1=1295$ points, from $(0,0,0,0)$ excluded to $(1,1,1,1)$ included, via $(0.2,0,0,0)$, etc. Unless otherwise noted, the output values of the evaluation functions were averaged over 5 simulations for each combination of the $\alpha_k$'s.

\subsection{Generation of urban patterns: \mbox{external} validation of the model}
\label{sec_extval}

\paragraph{Typical patterns}

We ran the model on different initial configurations, in which a few city centers $C_0$ (typically 4) were randomly positioned on a $56\times 56$ lattice, and their activity values drawn in $[1,a_{\max}]$ (both uniformly). The initial network $G_0$ was built progressively and quasi deterministically over increasing distances, by creating isolated clusters and linking them until they percolated into one component. The initial grid was empty ($\delta=0$ everywhere). Simulations were cut off at 30 iterations ($T=30\tau$), before the sprawl of urban settlements reached the square boundaries of the world and started ``reverberating''. Since this artifact occurred the fastest in a density-driven model, $\alpha_k=(1,0,0,0)$, we empirically assessed $T$ in that case and applied it everywhere.

\begin{figure}[t]
\centering
\includegraphics[trim=0mm 18mm 2mm 12mm, clip, width=\columnwidth]{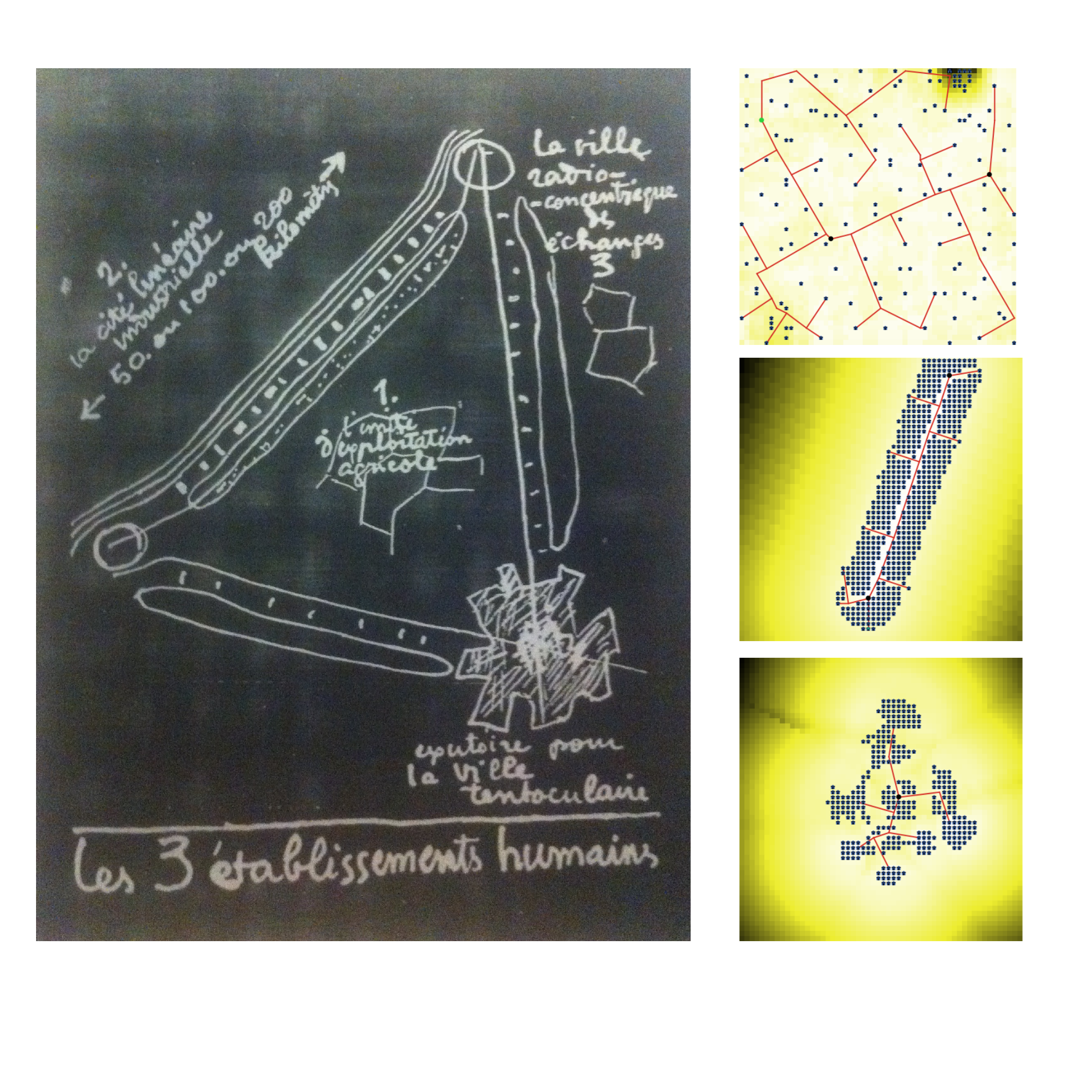}
\caption{\small Typical patterns obtained from our model, reproducing Le Corbusier's analysis of ``human settlements''. In his 1945 attempt to theorize urban planning, Le Corbusier analyzed the form of cities by hand and outlined three types of settlements: radial-concentric cities, linear cities along communication roads, and rural communities. We were able to reproduce this typology by setting the weights of the explicative variables of our model to corner values: \emph{Top-right}: $(\alpha_k)=(1,0,0,0)$, i.e.~density-based only. \emph{Middle}: $(0,1,0,0)$, i.e.~distance-to-road only. \emph{Bottom}: $(0.2,0,1,0)$, i.e.~network-distance combined with a little density. \emph{Left}: source~\cite{mangin2004ville}.}
\label{fig_corbu}
\end{figure}

Different parameter settings generated very diverse structures. In particular, we observed striking similarities between the patterns obtained for binary values of $\alpha_k$'s in some ``corners'' of the hypercube (one or two measures $d_k$ with weight 1, the others 0), and the fundamental urban configurations that Le Corbusier had identified in his 1945 analysis of human settlements~\cite{mangin2004ville} (Fig.~\ref{fig_corbu}).

\begin{figure}[htp]
\centering
\includegraphics[trim=0mm 5mm 0mm 23mm, clip, width=\columnwidth]{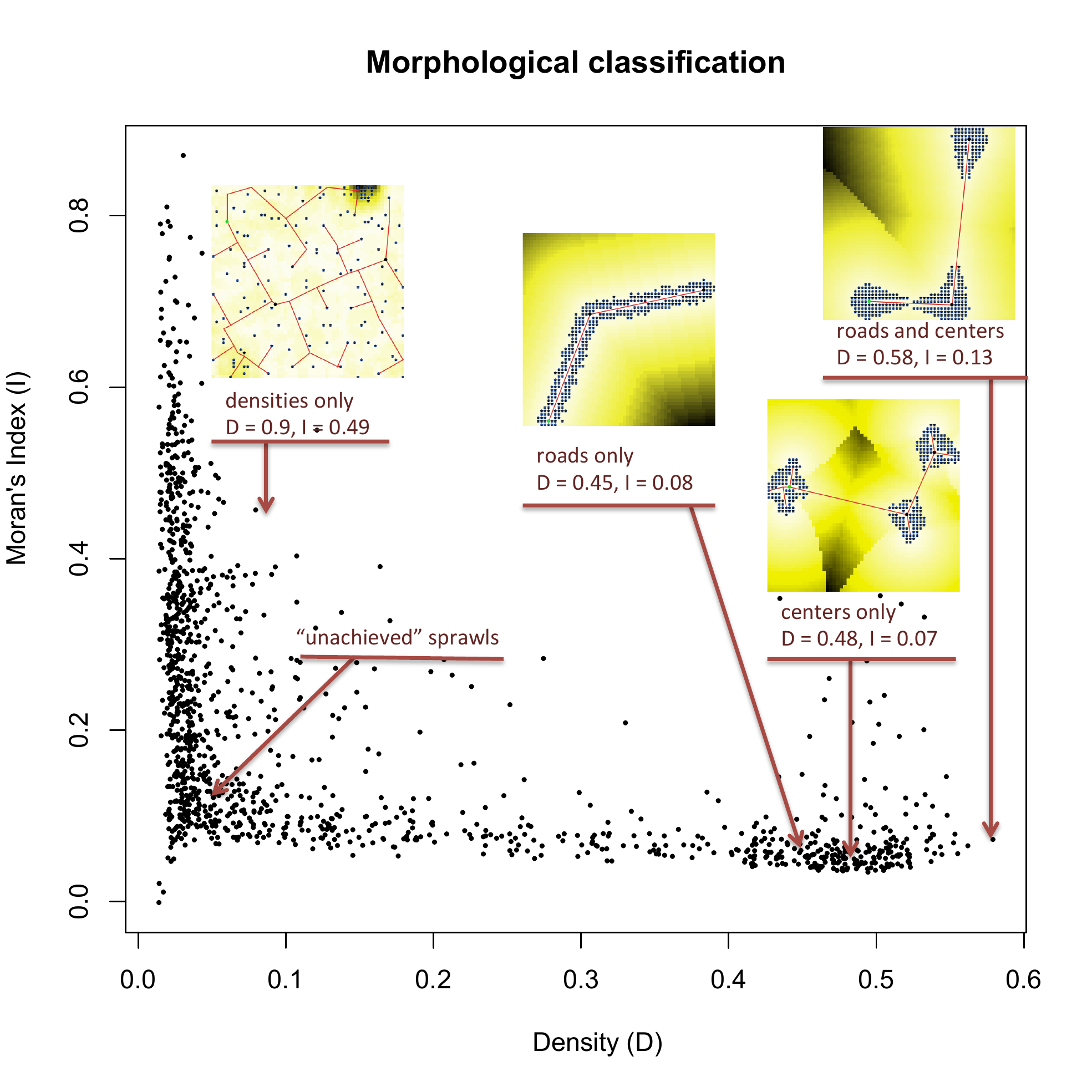}
\caption{\small Morphological classification of urban patterns. Scatterplot in the $(D,I)$ evaluation plane with four typical structures highlighted. Three of them, despite visual differences, are relatively close to each other in this space, indicating that two metrics are not sufficient for a full classification. The area near the origin corresponds to unfinished patterns, i.e.~which occupy only a small part of the world and cannot be compared with larger ones. Almost all $I$ values were positive.}
\label{fig_morpho}
\vspace{0.3cm}
\includegraphics[width=\columnwidth]{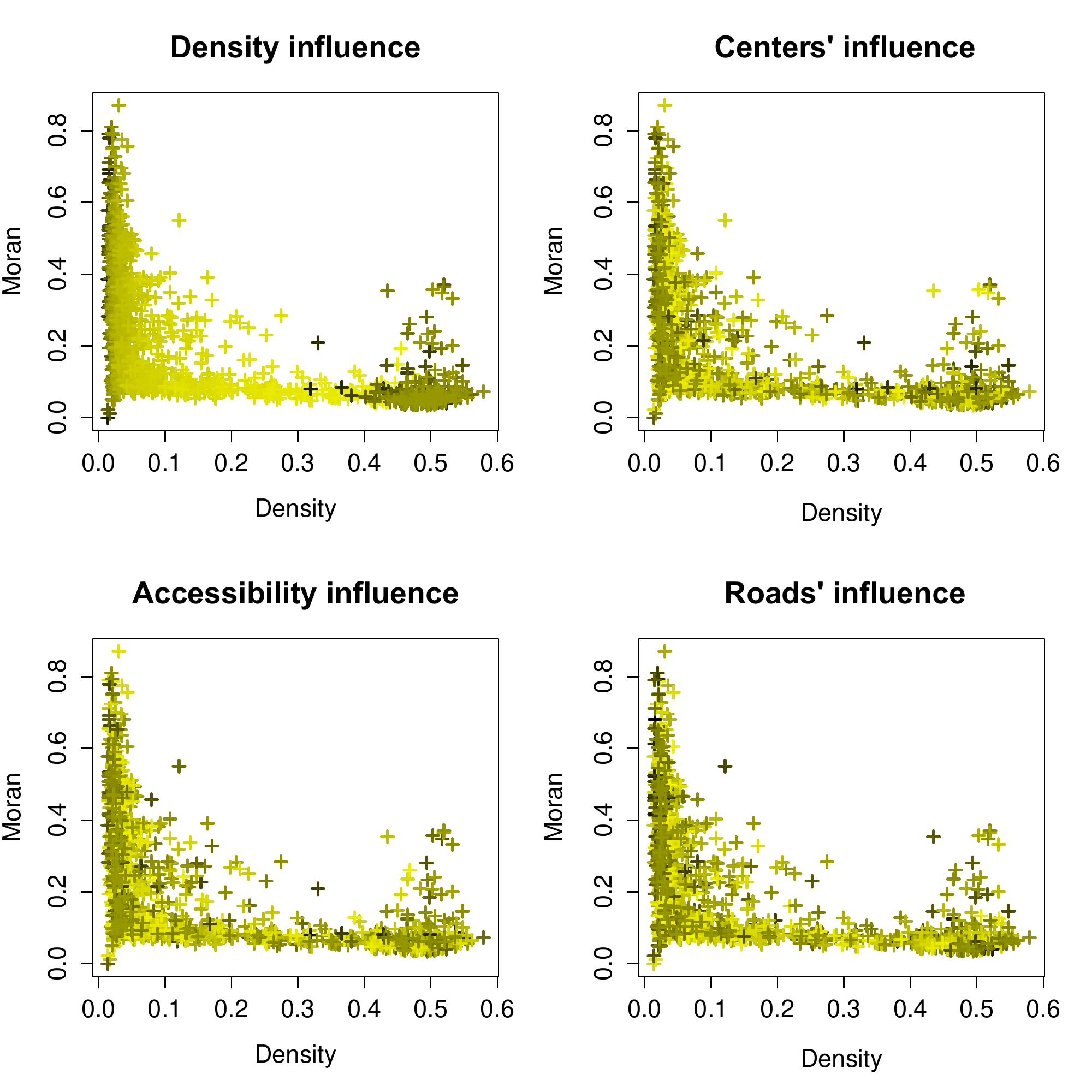}
\caption{\small Influence of each explicative variable $d_k$ on urban morphogenesis. Color darkness corresponds to the relative value of weight $\alpha_k$ used during the growth of mapped structures. Whereas Figs.~\ref{fig_morpho}-\ref{fig_influences} showed distinct classes at expected locations, this plot displays a rather uniform and chaotic distribution of high weights for $d_2$, $d_3$, and $d_4$, revealing a pervasive role of roads, city centers, and accessibility. Only density $d_1$ correlates better with its own evaluation function $D$ (a high influence of density results in low-density patterns), except for the \mbox{low-$I$} cluster on the right.}
\label{fig_influences}
\end{figure}

\paragraph{Classification of structures}

Using the pair of morphological indicators $(D,I)$ defined above, and by varying the $\alpha_k$'s, we constructed a 2D map of the dynamical regimes of our system (Figs.~\ref{fig_morpho}-\ref{fig_influences}), in which qualitatively different morphological ``classes'' could be distinguished. The projected locations of urban configurations in this plane allowed a better understanding and comparison of their features and growth process. Again, for certain corner parameter values (all of them 0 except one or two at 1), the results ended up in distinct locations on the map, which could be relatively well explained. Intermediate combinations of parameters, however, seemed to project the structures quite literally ``all over the map'', which might be interpreted as the emergence of chaos in the system.

\begin{figure}[t]
\centering
\includegraphics[trim=0mm 0mm 0mm 20mm, clip, width=\columnwidth]{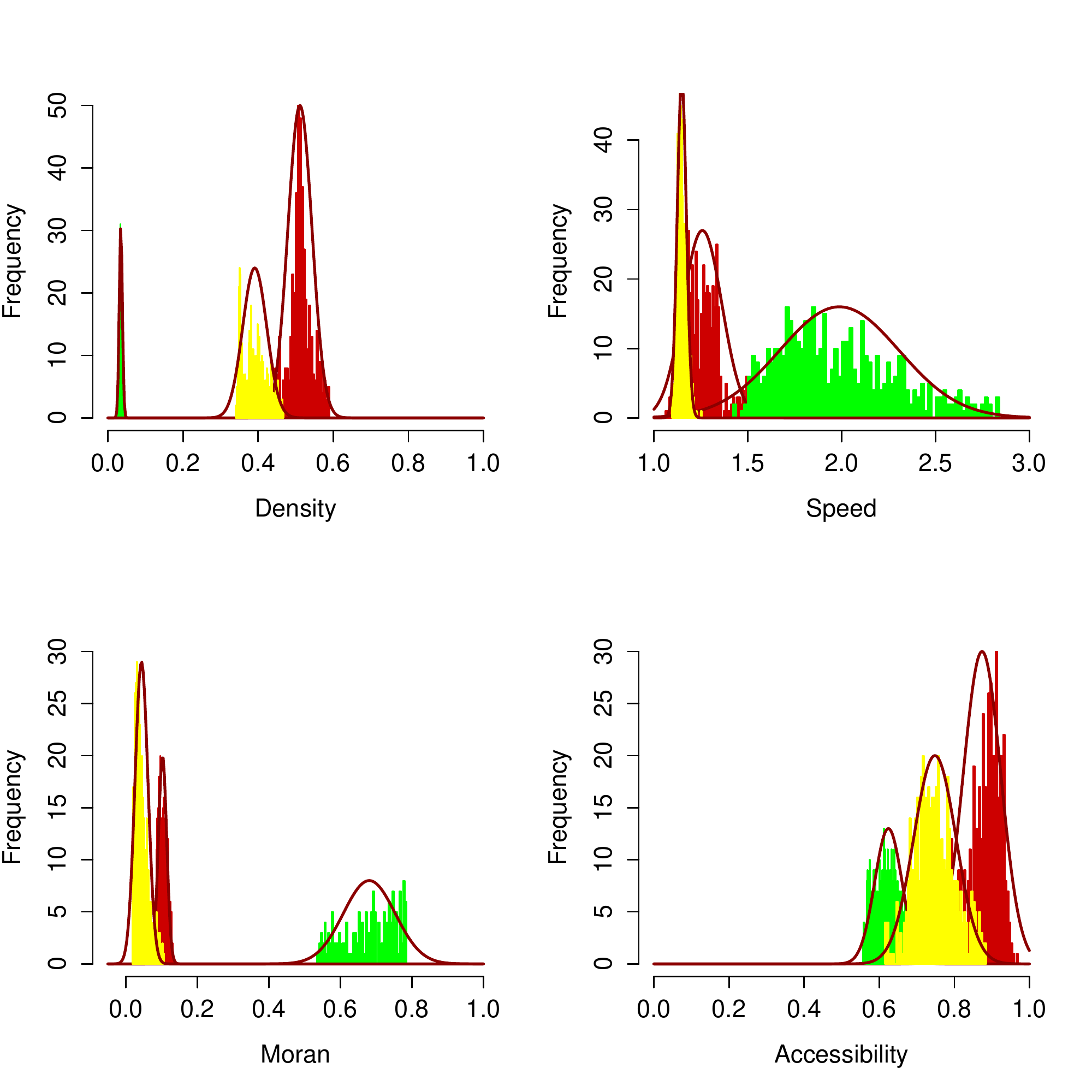}
\caption{\small Statistical distribution of the output evaluations. For each of the 15 corner points of the 4D hypercube of $\alpha_k$'s (excluding the origin), we ran 500 simulations from random initializations of 4 city centers $C_0$. Three resulting distributions out of these 15 are displayed, each in the form of a histogram of evaluation function values, $D$, $S$, $I$, and $A$, fitted with a Gaussian curve. \emph{Green}: $(\alpha_k)=(1,0,0,0)$, i.e.~a simulation taking into account only the density $d_1$. \emph{Yellow}: $(0,1,0,0)$, i.e.~Euclidean distance  $d_2$ only. \emph{Red}: $(0,0,0,1)$, i.e.~accessibility $d_4$ only. These three histograms were chosen for their minimum overlap and clarity of display; the other 17 are similar. The narrow peaks (except one), spread about the mean by $\pm10\%$, attest to the low sensitivity of the model with respect to the spatial initialization, and validates its internal consistency. This also allowed us to rely on a smaller number of runs in our experiments.}
\label{fig_hists}
\end{figure}

\begin{figure}[t]
\centering
\includegraphics[trim=0mm 0mm 5mm 24mm, clip, width=\columnwidth]{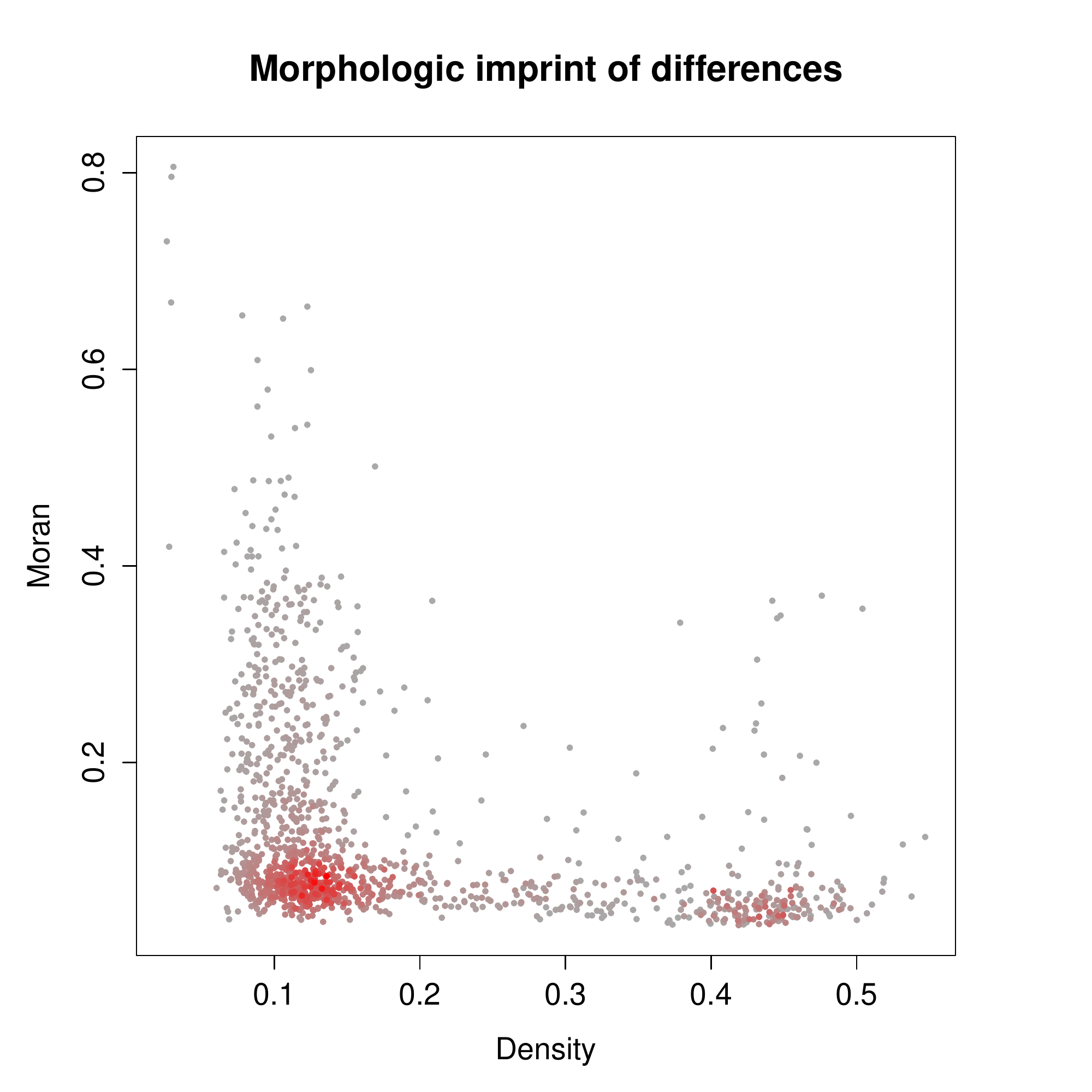}
\caption{\small Assessing the influence of the update scheme on the morphologies. In the $(D,I)$ classification plane, each point corresponds to 3 runs of a given combination of $\alpha_k$ parameters, repeated under a sequential ($n=1$) and under a parallel ($n=20$) update scheme. For each run, the symmetric difference $\Delta$ between the two patterns is computed and its average over the 3 runs is projected on the map. The color of a point highlights its ``significance'', defined as the product of its local density (clustered points represent more frequent configurations) and its pattern size, $|\Delta|$ (large patterns are more significant). The scattered points indicate that the model is sensitive to the update scheme for certain parameters. On the other hand, the concentration of significant points near the origin and $D=0.5$ means that corner cases, such as $(1,0,0,0)$, are more robust.}
\label{fig_imprint}
\end{figure}

\subsection{Sensitivity analysis and parameter space exploration: internal validation}

\paragraph{Sensitivity to initial conditions}

To ensure the validity of the results, we investigated the sensitivity of the model to the spatial conditions, the initial set of nodes $C_0$, estimating in particular the number of repetitions necessary to obtain statistically meaningful values for the evaluation functions. If conclusions drawn from one case were highly susceptible to small changes in the initial layout, then the model would obviously have less significance than if there was some invariance with respect to abstract topological features (in particular the distribution of centers' activities). The optimization heuristics would have to be designed very differently in these two cases.

Toward this assessment, we ran a large number of simulations under the same parameter values but starting from different initial $C_0$ configurations, and collected statistics on the output. For each of the 15 binary combinations of $\alpha_k$'s (excluding all zero), standard deviations were calculated over 500 runs. We obtained narrow peak distributions in most cases, with Gaussian widths typically at 10\% of the mean function value (Fig.~\ref{fig_hists}). In order to ensure that these were the typical widths on all the parameter space and not only on extreme binary points, we also explored the grid $\{0;0.5;1\}^{4}$ with 100 runs per point, assessing in a more representative subspace the relative spread of distributions. This confirmed that the evaluation functions were significantly less sensitive to the exact spatial locations than the parameters and overall topology, and justified our use of a smaller number of trials in subsequent experiments. Typically, assuming a normal distribution of width $\sigma = 0.1$, we needed $n=(2\sigma\!\cdot\!1.96/0.05)^2\simeq60$ trials to reach a 95\% confidence interval of length 0.05, and 5 trials for a length 0.17. For practical reasons of computing speed, we chose the latter.

\paragraph{Sensitivity to update scheme}

On the other hand, the emergent urban patterns depended on the number $n$ of cells filled at every iteration, before land values were recalculated at the next iteration, i.e.~whether the update scheme was a sequential ($n=1$) or parallel ($n>1$). Building several houses ``simultaneously'' between two market reevaluations is consistent with the view that real-world functions have a response delay, here of the order of $\tau$. There must be a limit, however, and an intermediate $n$ must be found to obtain reasonable simulations.

To this aim, we explored the 4D parameter space of the $\alpha_k$'s as in Figs.~\ref{fig_morpho}-\ref{fig_influences} and ran one sequential update scheme and one parallel update scheme with $n=20$ in each case. At the end of the simulation, $t=T$, the two corresponding output patterns $\delta_\mathrm{seq}$ and $\delta_\mathrm{par}$ were compared by calculating their symmetric difference, i.e.~the subset of lattice cells that were built either in one or the other but not in both: $\Delta = \{(i,j);\;\delta_\mathrm{seq}(i,j,T) \neq \delta_\mathrm{par}(i,j,T)\}$. Then, these difference patterns $\Delta$ were projected on the same classification map $(D,I)$ used previously (Fig.~\ref{fig_imprint}). The results showed that for many combinations of parameters, the model's behavior could be noticeably influenced by the update scheme, as many difference patterns exhibited a nontrivial structure with high density or high Moran's index or both. On the other hand, it exhibited a stronger invariance for the corner quadruplets of $\alpha_k$'s: in these cases the $\Delta$'s clustered near the origin and $D=0.5$. Based on this study, we decided to adopt a parallel update scheme with $n=15$ built cells per time step in the remainder of the experiments.

\paragraph{Exploration of parameter space}

The above two preliminary studies validated the robustness of the model with respect to the initialization and update scheme, and helped us choose a reasonable number of runs (about 5) for each parameter combination, and a adequate degree of parallelism in the simulations ($n=15$). Next, we revisited the $\alpha_k$ hypercube (same 1295 points in the partition of step 0.2), this time calculating the complete charts of all evaluation functions. Other parameters with a direct correspondence to the real-world, depending on the scale adopted, were set to fixed values. For example, the neighborhood radius $\rho$ or the road-triggering distance $\theta_2$ were both equal to 5 cells: this number could represent 50m, characteristic of a block at the scale of a district, or 500m for a district in a city, or 5km between cities in a region. 

\begin{figure}[t]
\centering
\includegraphics[trim=0mm 48mm 0mm 42mm, clip, width=\columnwidth]{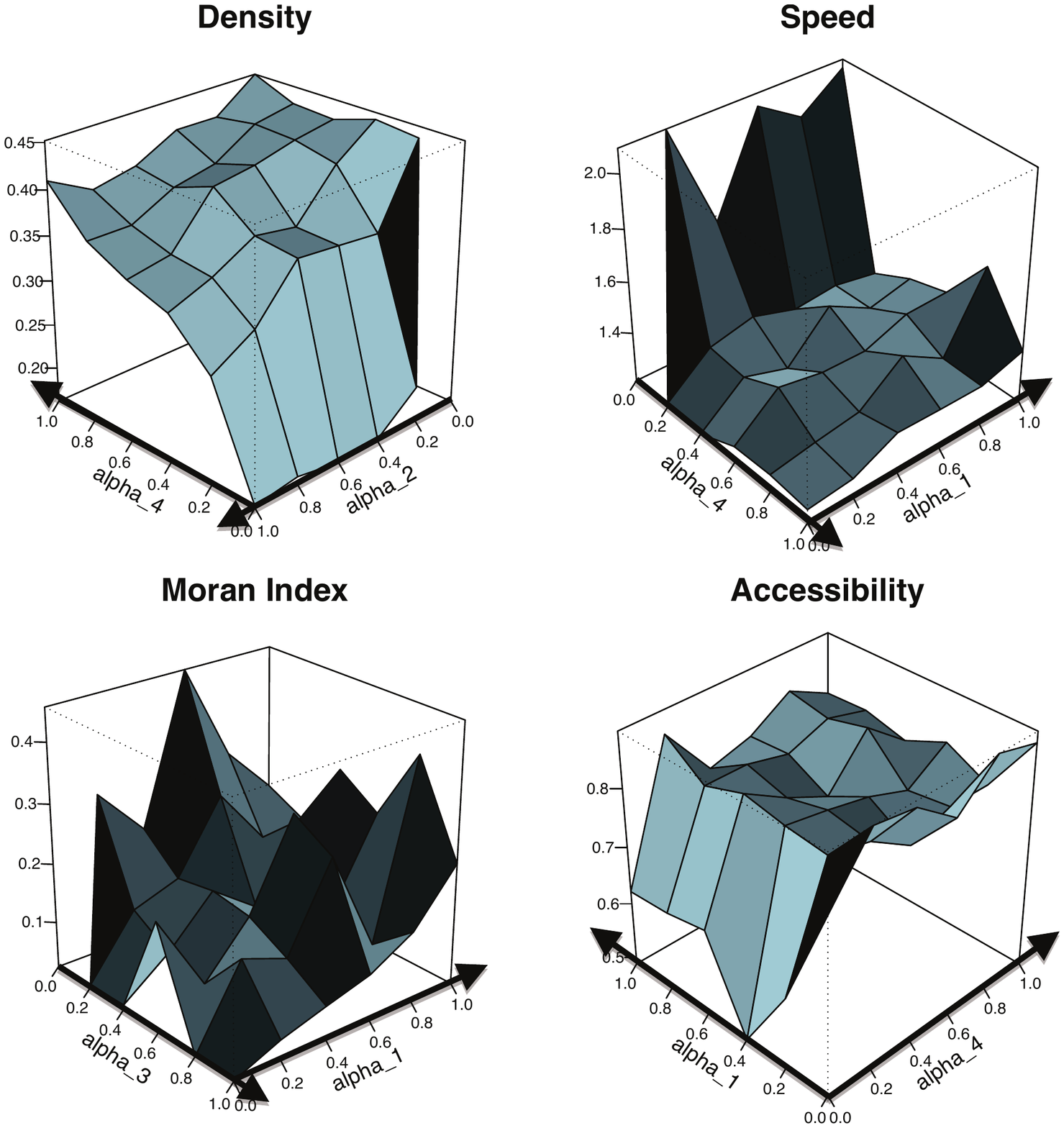}
\caption{\small Sample surface plots of the evaluation functions. For each 4D field of evaluation values in the hypercube, we select two out of four parameters and display the 2D slice corresponding to the other two parameters set to $(0,0)$. Horizontal axes are reoriented in each case to minimize visual clutter. This exhaustive exploration has an intrinsic explanatory value (see text), and allows us to predict with some level of confidence how the model responds to certain input parameters.}
\label{fig_plots3d}
\end{figure}

Examples of evaluation surfaces in 2D projection spaces are shown in Fig.~\ref{fig_plots3d}. Each function, $D$, $I$, $S$, and $A$, was plotted against two parameters out of four, chosen for their higher ``influence'' (variations in amplitude) on the function. The economic index $H$ was not calculated here (see~\ref{sec_practapp}). This exhaustive exploration of parameter space was necessary to gain deeper insight into the behavior of the model. It also represents a crucial step toward making computational simulations more rigorous~\cite{banos2013HDR}.

Altogether, we observed that outputs varied for the most part smoothly, except Moran's index which appeared more chaotic. Variations were greater in cases where one parameter was dominant. For example, the measures of density $D$, speed $S$ and (global) accessibility $A$ all exhibited a significant jump when including the effect of (local) accessibility $d_4$ in the simulations, i.e.~when transitioning from $\alpha_4=0$ to $\alpha_4>0$. In particular, the more activities were influent, the denser the city became---a nonintuitive emergent effect, compared to top-down planning alternatives that would try to optimize accessibility while keeping density low. Speed, or rather ``sluggishness'', exploded when density was the only influence on urban sprawl: this confirmed that pure density-driven dynamics creates anarchic growth, without concern for network performance.

As for global accessibility, or rather the difficulty thereof, it was minimal for $\alpha_4=0$ : an interesting paradoxical effect suggesting that when individual agents took into account local accessibility ($\alpha_4>0$), a few of them might have occupied the ``best spots'' too quickly, significantly diminishing the others' prospects. Therefore, at the collective level, it would be better for everyone to ignore that dimension---an example where competition at the individual level does not produce the most efficient system for all. Finally, no meaningful conclusion could be formulated about the chaotic variations of Moran's index, except for its extreme sensitivity to spatial structure.

\begin{figure*}[htp]
\centering
\includegraphics[height=0.45\textwidth]{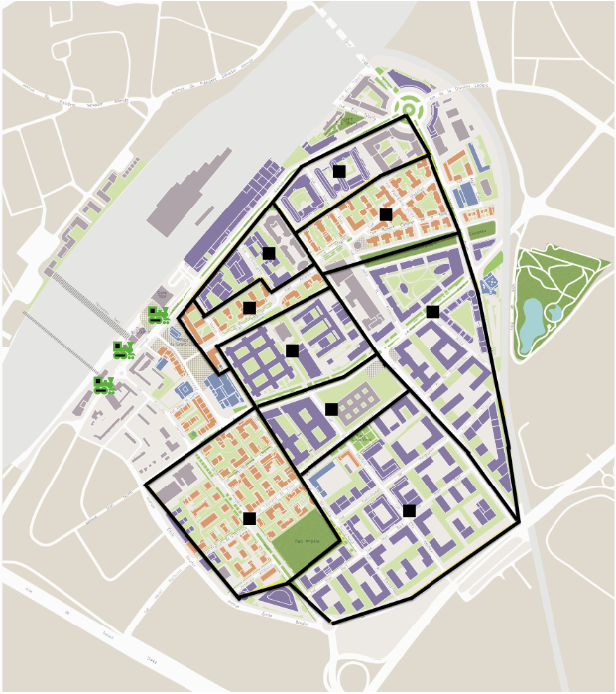}\hspace{0.5cm}\includegraphics[height=0.45\textwidth]{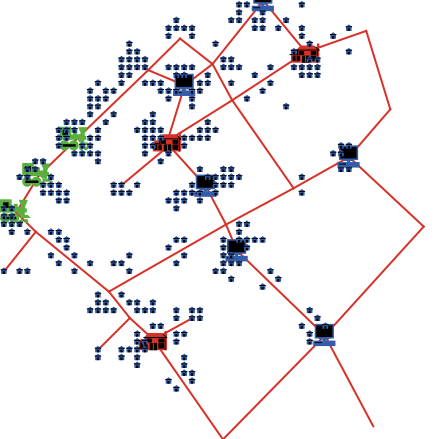}
\caption{\small Practical application: optimizing the distribution of activities over urban centers. \emph{Left}: existing masterplan of the Massy Atlantis district: 9 building areas are identified by their centers (black squares). Buildings are color-coded by type of activity (orange: residential, blue: tertiary). Railroad tracks traverse the upper left corner (light blue strip) and three train transportation hubs are also represented (green locomotive icons): these are part of the fixed environment. \emph{Right}: example of housing configuration obtained after a simulation in which the 9 centers' activities were initialized to the same values as their real-world counterpart (red house icons: residential, blue computer icons: tertiary; one of 510 possible distributions). The evaluation functions of the outcome are $(H,A)=(0.067,0.76)$, a point close to the Pareto front of Fig.~\ref{fig_pareto} (blue circle).}
\label{fig_atlantis}
\end{figure*}

\subsection{Practical example}
\label{sec_practapp}

In this section, we apply our model to the optimization of activities on top of a real-world urban structure obtained from a geographic file, as opposed to an randomly generated, artificial configuration. This type of scenario occurs in a planning problem where one must decide about the possible land use of predefined zones.

The practical example under study here concerns the planning of a new district. It is based on a real-world neighborhood, Massy Atlantis (Paris metropolitan area), built in 2012 (Fig.~\ref{fig_atlantis}). We would like to investigate whether a more efficient planning could have been achieved. The goal of this exercise is to find an optimal assignment of two types of activities, ``residential'' (apartments) or ``tertiary'' (offices), to the centers of 9 areas located on a map. The transportation infrastructure is already in place and the train station is also considered a center with a fixed, third type of activity. A network of avenues is laid out and passes through the 9 centers. The district is initially empty (unbuilt). The particular spatial configuration was automatically imported from a GIS shapefile, so the computation could be readily applied to other cases.

Parameters of the model were set as follows: high influence of activities, $\alpha_4=1$, reflecting the fact that accessibility to home, workplace, and train station are of special importance to the agents of this district; medium influence of density, $\alpha_1=0.7$, because, not far from Paris, housing must reasonably fill the available areas; no influence of road proximity, $\alpha_2=0$, since the initial network is already built and the scale is relatively small; and no influence of network-distance, $\alpha_3=0$, because centers in this problem are abstract entities representing areas. 

For every possible distribution of binary activities over the 9 areas, excluding the two uniform cases (all residential or all tertiary), the model was simulated 5 times, producing a total of $(2^9-2)\times 5=2550$ runs. The resulting configurations were examined here via a morphological projection in the $(H,A)$ plane, instead of $(D,I)$ used in the previous sections, as we judged it to be a more meaningful measure of fitness in this application. The calculation of the economic segregation index $H$ involved a secondary agent-based simulation on top of the main urban development model (details not provided here).

\begin{figure}[t]
\centering
\includegraphics[trim=0mm 0mm 10mm 10mm, clip, width=\columnwidth]{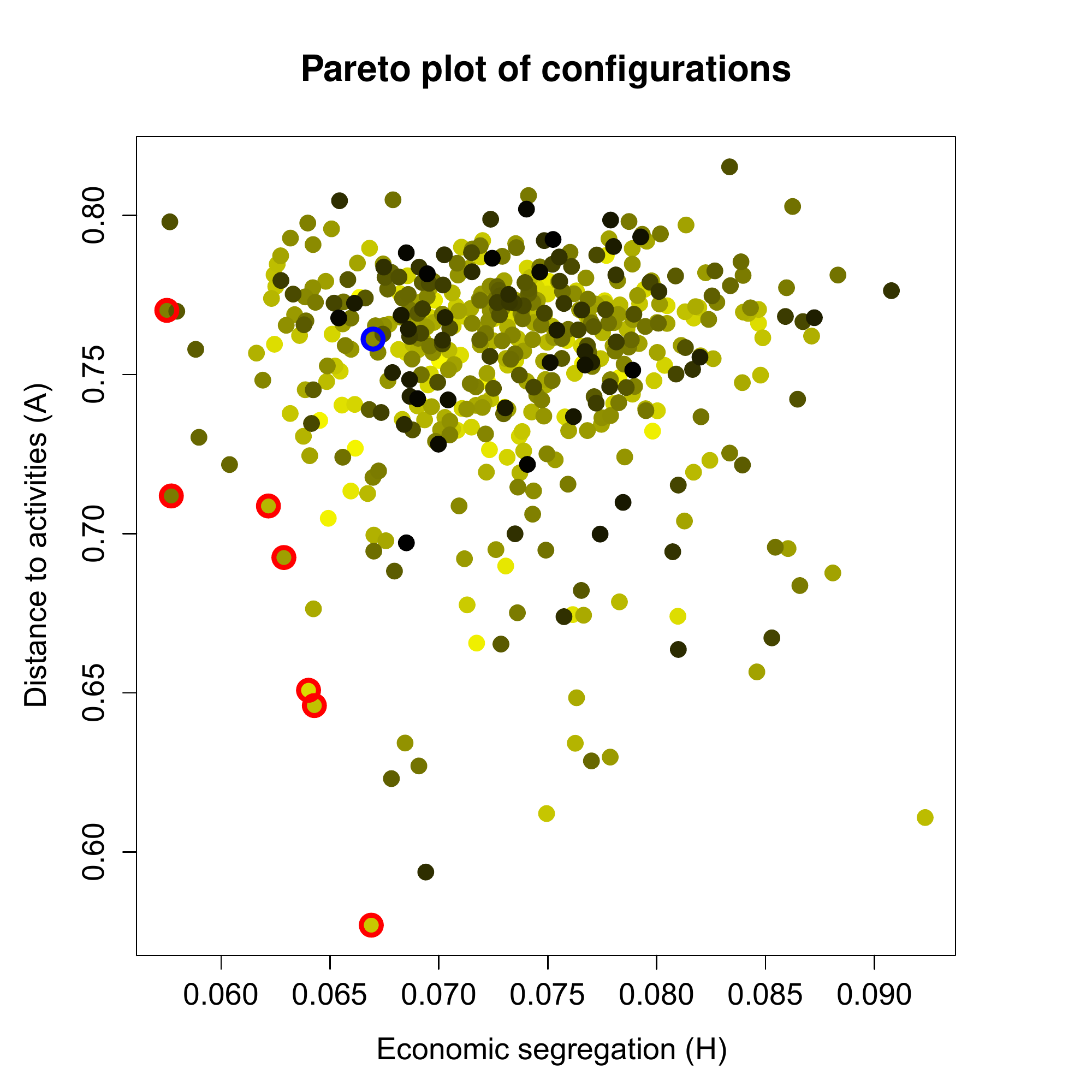}
\caption{\small Scatterplot of all configurations in the $(H,A)$ morphological plane. A Pareto front (red circles) is apparent in the bottom left part of the plot: it corresponds to ``optimal'' configurations trying to minimize both $H$ and $A$ objectives. The real situation (blue circle in $H=0.067, A=0.76$) corresponds to Fig.~\ref{fig_atlantis} and is not far from this front. Points are colored according to their level of heterogeneity $\lambda$, from low (black) to high (yellow). More homogeneous configurations are concentrated in the central cluster, whereas Pareto points and their neighbors have higher heterogeneity levels. This lends support to the principle of ``functional diversity'', which is often adopted by planners and urbanists today but has never been backed up by computational simulations.}
\label{fig_pareto}
\end{figure}

Results are shown in Fig.~\ref{fig_pareto}. We obtained a Pareto front of ``optimal solutions'' trying to minimize both $H$ and $A$, while observing that the actual configuration is not far from being optimal itself, and appears to be a compromise between accessibility and economic performance. After closer examination of the Pareto front and its vicinity, we found that the distribution of activities was highly mixed in these points. More precisely, we defined a spatial heterogeneity index of center activities by
\begin{equation}
\lambda=a_\mathrm{max}\frac{\sum_{\scriptstyle c\neq c'\atop\scriptstyle a(c)\neq a(c')}d(c,c')^{-1}}{\sum_{c\neq c'}d(c,c')^{-1}}
\end{equation}
where $c=(i,j)$ and $c'=(i',j')$ are two centers, $d(c, c')$ their Euclidean distance, and $a(c)$, $a(c')$ their activities. Points in the scatterplot were colored according to their level of $\lambda$. Highly heterogeneous configurations appeared in regions of the plot distinct from homogeneous configurations, which were for the most part located in the central cluster. Optimal solutions and their neighbors all corresponded to high heterogeneity. This interesting result is a step toward evidence-based justification of mixed land use in planning---a principle often invoked by urbanists but never quantitatively demonstrated.

In conclusion, this case study is encouraging as it proposes a concrete methodology of optimal planning with respect to criteria that are relevant to a particular situation. It could be used by generically planners in similar situations, while remaining cautious on the conditions of its applicability. We discuss this point next.

\section{Discussion} \label{sec_discussion}

The reproduction of typical urban morphologies and the possible application to a real-world problem shown in the previous sections indicate that a model like ours can be useful for evidence-based decision-making in urban planning. Several questions remain open, however, and would need further investigation.

\paragraph{Scale of the model}

One ambiguity of the model is that it can be applied at different scales, therefore there is no unique correspondence between its agents and the real world. As the above results illustrate, the simulated urban configurations may represent a system of cities at the macroscopic scale, the neighborhoods of one city at the mesoscopic scale, or the buildings of one district at the microscopic scale. Without engaging in an ontological debate over levels of abstraction, this could still be pointed out as a potential issue.

We wish to argue, however, that the multiscale applicability of our model is legitimate as a great number of urban systems and associated dynamics have been shown to be ``scale-free'', in particular by Pumain~\cite{pumain2004scaling}, and even to possess fractal properties, by Batty~\cite{FractalCities}. It means that scaling laws may also operate in our model, therefore qualitative results should remained unchanged while the quantitative evolution of variables and relations should only depend on the underlying power law's exponent.

Barth\'elemy~\cite{2013arXiv1309.3961L} warns that most multi-agent urban models fail because they do not focus on the ``dominating'' physical process but, instead, integrate too many aspects that bear no relevance to the emergent properties of the system. Following up on this advice, we believe that we have successfully identified ``good'' proxies for the dominating processes of urban morphogenesis, namely: urban density, accessibility to road network, and accessibility to main functionalities.

\paragraph{Local scope}

When the model is considered at a mesoscopic or microscopic scale, another objection could be that it seems to limit itself to an artificially ``closed'' urban system, neglecting important contextual phenomena such as economic exchanges. Yet, although input and output flows are greatly simplified here, they are still present in implicit form. Our simulated world is not truly closed, since newly built houses are associated with a net influx of resources. Moreover, despite the absence of a direct economic force in the growth dynamics (the $H$ index is only a post-hoc metric), the attractivity of centers constitutes a proxy for underlying activity, and a form of interdependence among urban processes. Finally, other models that have taken into account the global complex network of cities~\cite{andersson2003urban} have reproduced well-known patterns of urban systems much like ours.

Therefore, here too, local or global approaches appear to be equivalent and the modeling decisions and compromises made in each case must be compared. This question also ties in with the fundamental issue, contained in the previous point, of the existence of a ``minimal dimension'' for a generalized representation of urban systems. The challenge is to understand how universal the dependence between a system and its dimension may be, and if a generalized minimalist formulation can be constructed. Speculations toward that ambitious goal have been formulated by Haken~\cite{haken2003face} through a notion of ``semantic information'' linked to properties of attractors in dynamical systems. This theory, however, has not been quantified, i.e.~neither confirmed nor falsified.

\paragraph{Quantitative calibration}

The question of the validity of the model is also linked to the need for a finer quantitative calibration based on real patterns, which creates a dilemma: on the one hand, calibration on the errors of output function proxies does not influence the formation of spatial patterns; on the other hand, calibration on the spatial patterns themselves is too constraining and may preclude the emergence of other, similar patterns. Previous works addressing the issue of calibration~\cite{maria2003stochastic} have not been conclusive so far.

To revisit this question, we would need to apply our model at a finer grain of spatial resolution, i.e.~a very large world in terms of data size. In this scenario, it would be particularly important to keep processing time under control by reducing computational complexity, for example through a cache of the network's shortest paths. The potential increase in size can also create methodological hurdles, not just computational, as a huge amount of details in the resulting patterns might contribute to more noise than signal and significantly bias the indicators. One solution would be to create a new operator extracting the morphological envelope of the generated pattern, along the lines of an original method proposed by Frankhauser et al.~\cite{frankhauser2005multi,tannier:halshs-00461657}. Other ways to deal with noise may involve Gaussian smoothing.

\paragraph{Complex coupling with economic model}

Our method of economic evaluation consists of ``simple coupling'', i.e. running a secondary agent-based model (the basis of $H$'s calculation, not described here) after the primary urban growth simulation has finished. Another important direction of research would implement a ``complex coupling'' between the two models in the sense proposed by Varenne~\cite{varenne2013modeliser}: the study of urban sprawl on other time scales would require the \emph{simultaneous} and mutually interacting evolutions of the population, the building rents, and the terrain values. Obviously, this would lead to a more sophisticated model oriented toward a whole new set questions, such as the evaluation of long-term rent policies to foster social diversity.

\section{Conclusion} \label{sec_conclusion}

We have proposed a hybrid network/grid model of urban growth structures, and studied their morphological and functional properties by simulation. Results showed that it could reproduce the characteristic urban facts of a classical typology of ``human settlements'', and was also applicable to a concrete scenario by calculating ``optimal'' solutions (in the Pareto sense) to a planning challenge in an existing zoning context. Our work provide evidence in favor of the ``mixed-use city'', a topic on which literature is still scarce and future work is needed. This paradigm is now commonly advocated by urbanists, such as Mangin~\cite{mangin2004ville} through his concept of \emph{``ville passante''} (a pun on ``evolving/flowing/pedestrian city''), and would require more validation through quantitative results.

Finally, beyond its technical achievements and potential usefulness as a decision-making tool, our work also fuels a contemporary debate on the state-of-the-art in ``quantitative urbanism''. Siding with Portugali~\cite{portugali2012complexity}, we certainly agree that the conception and application of computational models is a delicate matter, which can lead to more confusion than explanation if not properly handled and validated. Depending on the scale, a careless choice of parameter values can produce dubious results. Yet, we support the idea that \emph{quantitative} insights are paramount for a better understanding of urban and social systems. With the recent explosion in data size and computing power, evidence-based analysis and solutions are becoming a real alternative to older attitudes, such as Lefebvre's~\cite{henri1968droit}, which doubted that scientific approaches could ever translate or predict the mechanisms of a city.

\footnotesize
\bibliographystyle{abbrv}

\begin{thebibliography}{10}

\bibitem{andersson2003urban}
C.~Andersson, A.~Hellervik, K.~Lindgren, A.~Hagson, and J.~Tornberg.
\newblock Urban economy as a scale-free network.
\newblock {\em Physical Review E}, 68(3):036124, 2003.

\bibitem{banos2012network}
A.~Banos.
\newblock Network effects in schelling's model of segregation: new evidences
  from agent-based simulation.
\newblock {\em Environment and Planning B: Planning and Design},
  39(2):393--405, 2012.

\bibitem{banos2013HDR}
A.~Banos.
\newblock {\em Pour des pratiques de mod{\'e}lisation et de simulation
  lib{\'e}r{\'e}es en G{\'e}ographie et SHS}.
\newblock PhD thesis, UMR CNRS G{\'e}ographie-Cit{\'e}s, ISCPIF, D{\'e}cembre
  2013.

\bibitem{banos2012towards}
A.~Banos and C.~Genre-Grandpierre.
\newblock Towards new metrics for urban road networks: Some preliminary
  evidence from agent-based simulations.
\newblock In {\em Agent-based models of geographical systems}, pages 627--641.
  Springer, 2012.

\bibitem{batty1997cellular}
M.~Batty.
\newblock Cellular automata and urban form: a primer.
\newblock {\em Journal of the American Planning Association}, 63(2):266--274,
  1997.

\bibitem{Bat07}
M.~Batty.
\newblock {\em Cities and Complexity: Understanding Cities with Cellular
  Automata, Agent-based Models, and Fractals}.
\newblock MIT Press, 2007.

\bibitem{batty2013new}
M.~Batty.
\newblock {\em The New Science of Cities}.
\newblock MIT Press, 2013.

\bibitem{FractalCities}
M.~Batty and P.~Longley.
\newblock {\em Fractal Cities: A Geometry of Form and Function}.
\newblock Academic Press, London, 1994.

\bibitem{batty1997possible}
M.~Batty and Y.~Xie.
\newblock Possible urban automata.
\newblock {\em Environment and Planning B}, 24:175--192, 1997.

\bibitem{benenson1998multi}
I.~Benenson.
\newblock Multi-agent simulations of residential dynamics in the city.
\newblock {\em Computers, Environment and Urban Systems}, 22(1):25--42, 1998.

\bibitem{DBM11}
G.~Caruso, G.~Vuidel, J.~Cavailhes, P.~Frankhauser, D.~Peeters, and I.~Thomas.
\newblock Morphological similarities between dbm and a microeconomic model of
  sprawl.
\newblock {\em Journal of Geographical Systems}, 13:31--48, 2011.

\bibitem{frankhauser2005multi}
P.~Frankhauser and C.~Tannier.
\newblock A multi-scale morphological approach for delimiting urban areas.
\newblock In {\em 9th Computers in Urban Planning and Urban Management
  conference (CUPUM'05), University College London}, 2005.

\bibitem{gauvin2009phase}
L.~Gauvin, J.~Vannimenus, and J.-P. Nadal.
\newblock Phase diagram of a schelling segregation model.
\newblock {\em The European Physical Journal B}, 70(2):293--304, 2009.

\bibitem{golden2012modeling}
B.~Golden, M.~Aiguier, and D.~Krob.
\newblock Modeling of complex systems ii: A minimalist and unified semantics
  for heterogeneous integrated systems.
\newblock {\em Applied Mathematics and Computation}, 218(16):8039--8055, 2012.

\bibitem{haken2003face}
H.~Haken and J.~Portugali.
\newblock The face of the city is its information.
\newblock {\em Journal of Environmental Psychology}, 23(4):385--408, 2003.

\bibitem{henri1968droit}
L.~Henri.
\newblock Le droit {\`a} la ville.
\newblock {\em Anthropos}, 1968.

\bibitem{heppenstall2012agent}
A.~J. Heppenstall, A.~T. Crooks, and L.~M. See.
\newblock {\em Agent-based models of geographical systems}.
\newblock Springer, 2012.

\bibitem{hillier1976space}
B.~Hillier, A.~Leaman, P.~Stansall, and M.~Bedford.
\newblock Space syntax.
\newblock {\em Environment and Planning B: Planning and Design}, 3(2):147--185,
  1976.

\bibitem{iltanen2012cellular}
S.~Iltanen.
\newblock Cellular automata in urban spatial modelling.
\newblock In {\em Agent-based models of geographical systems}, pages 69--84.
  Springer, 2012.

\bibitem{lenechet:hal-00696445}
F.~Le~N{\'e}chet and A.~Aguilera.
\newblock D{\'e}terminants spatiaux et sociaux de la mobilit{\'e}
  domicile-travail dans 13 aires urbains fran{\c c}aises : une approche par la
  forme urbaine, {\`a} deux {\'e}chelles g{\'e}ographiques.
\newblock In {\em {ASRDLF 2011}}, SCHOELCHER, Martinique, July 2011.
\newblock http://asrdlf2011.com/.

\bibitem{2013arXiv1309.3961L}
R.~{Louf} and M.~{Barthelemy}.
\newblock {Modeling the polycentric transition of cities}.
\newblock {\em ArXiv e-prints}, Sept. 2013.

\bibitem{mangin2004ville}
D.~Mangin.
\newblock {\em La ville franchis{\'e}e: formes et structures de la ville
  contemporaine}.
\newblock {\'E}ditions de la Villette Paris, 2004.

\bibitem{maria2003stochastic}
C.~Maria~de Almeida, M.~Batty, A.~M. Vieira~Monteiro, G.~C{\^a}mara, B.~S.
  Soares-Filho, G.~C. Cerqueira, and C.~L. Pennachin.
\newblock Stochastic cellular automata modeling of urban land use dynamics:
  empirical development and estimation.
\newblock {\em Computers, Environment and Urban Systems}, 27(5):481--509, 2003.

\bibitem{MBB09}
D.~Moreno, D.~Badariotti, and A.~Banos.
\newblock Un automate cellulaire pour exp{\'e}rimenter les effets de la
  proximit{\'e} dans le processus d'{\'e}talement urbain : le mod{\`e}le
  raumulus.
\newblock {\em Cybergeo : European Journal of Geography}, 2009.

\bibitem{moreno2007conception}
D.~Moreno, A.~Banos, and D.~Badariotti.
\newblock Conception d'un automate cellulaire non stationnaire {\`a} base de
  graphe pour mod{\'e}liser la structure spatiale urbaine: le mod{\`e}le remus.
\newblock {\em Cybergeo: European Journal of Geography}, 2007.

\bibitem{peeters2009space}
D.~Peeters and M.~Rounsevell.
\newblock Space time patterns of urban sprawl, a 1d cellular automata and
  microeconomic approach.
\newblock {\em Environment and Planning B: Planning and Design}, 36:968--988,
  2009.

\bibitem{portugali2012complexity}
J.~Portugali.
\newblock Complexity theories of cities: Achievements, criticism and
  potentials.
\newblock In {\em Complexity Theories of Cities Have Come of Age}, pages
  47--62. Springer, 2012.

\bibitem{portugali2012book}
J.~Portugali, H.~Meyer, E.~Stolk, and E.~Tan.
\newblock {\em Complexity theories of cities have come of age: an overview with
  implications to urban planning and design}.
\newblock Springer, 2012.

\bibitem{pumain2004scaling}
D.~Pumain.
\newblock Scaling laws and urban systems.
\newblock {\em Santa Fe Institute, Working Paper n 04-02}, 2:26, 2004.

\bibitem{pumain2012multi}
D.~Pumain.
\newblock Multi-agent system modelling for urban systems: The series of simpop
  models.
\newblock In {\em Agent-based models of geographical systems}, pages 721--738.
  Springer, 2012.

\bibitem{QGIS_software}
{QGIS Development Team}.
\newblock {\em QGIS Geographic Information System}.
\newblock Open Source Geospatial Foundation, 2009.

\bibitem{R}
{R Core Team}.
\newblock {\em R: A Language and Environment for Statistical Computing}.
\newblock R Foundation for Statistical Computing, Vienna, Austria, 2013.

\bibitem{schelling1969models}
T.~C. Schelling.
\newblock Models of segregation.
\newblock {\em The American Economic Review}, 59(2):488--493, 1969.

\bibitem{tannier:halshs-00461657}
C.~Tannier, G.~Vuidel, and P.~Frankhauser.
\newblock {D{\'e}limitation d'ensembles morphologiques par une approche
  multi-{\'e}chelle}.
\newblock In J.-C. Folt{\^e}te, editor, {\em {Actes des huiti{\`e}mes
  Rencontres de Th{\'e}o Quant}}, page~14, Besan{\c c}on, France, 2008.
\newblock http://thema.univ-fcomte.fr/theoq/.

\bibitem{tsai2005quantifying}
Y.-H. Tsai.
\newblock Quantifying urban form: compactness versus' sprawl'.
\newblock {\em Urban Studies}, 42(1):141--161, 2005.

\bibitem{van2012activity}
J.~van Vliet, J.~Hurkens, R.~White, and H.~van Delden.
\newblock An activity-based cellular automaton model to simulate land-use
  dynamics.
\newblock {\em Environment and Planning-Part B}, 39(2):198, 2012.

\bibitem{varenne2013modeliser}
F.~Varenne, M.~Silberstein, et~al.
\newblock {\em Mod{\'e}liser \& simuler. Epist{\'e}mologies et pratiques de la
  mod{\'e}lisation et de la simulation, tome 1}.
\newblock 2013.

\bibitem{white2006modeling}
R.~White.
\newblock Modeling multi-scale processes in a cellular automata framework.
\newblock In {\em Complex artificial environments}, pages 165--177. Springer,
  2006.

\bibitem{white1993cellular}
R.~White and G.~Engelen.
\newblock Cellular automata and fractal urban form: a cellular modelling
  approach to the evolution of urban land-use patterns.
\newblock {\em Environment and planning A}, 25(8):1175--1199, 1993.

\bibitem{NetLogo}
U.~Wilensky.
\newblock Netlogo.
\newblock Center for Connected Learning and Computer-Based Modeling,
  Northwestern University, Evanston, IL., 1999.

\bibitem{Wu96alinguistic}
F.~Wu.
\newblock A linguistic cellular automata simulation approach for sustainable
  land development in a fast growing region.
\newblock {\em Computers, Environment and Urban Systems}, 20:367--87, 1996.

\end{thebibliography}

\end{document}